\documentclass[manuscript]{aastex6}

\usepackage{graphicx}
\usepackage{natbib}
\usepackage[english]{babel}  
\usepackage{subfigure}
\usepackage{url}
\newcommand{\aftr}{  }

\begin{document}

\title{Plasma sloshing in pulse-heated solar and stellar coronal loops}



\author{F. Reale\altaffilmark{1}}
\affil{Dipartimento di Fisica \& Chimica, Universit\`a di Palermo,
              Piazza del Parlamento 1, 90134 Palermo, Italy;
              fabio.reale@unipa.it}

\altaffiltext{1}{INAF-Osservatorio Astronomico di Palermo, Piazza del Parlamento 1, 90134 Palermo, Italy}

\begin{abstract}
There is evidence that coronal heating is highly intermittent, and flares are the high energy extreme. The properties of the heat pulses are difficult to constrain. Here hydrodynamic loop modeling shows that several large amplitude oscillations ($\sim 20$\% in density) are triggered in flare light curves if the duration of the heat pulse is shorter that the sound crossing time of the flaring loop. The reason is that the plasma has not enough time to reach pressure equilibrium during the heating and traveling pressure fronts develop. The period is a few minutes for typical solar coronal loops, dictated by the sound crossing time in the decay phase. The long period and large amplitude make these oscillations different from typical MHD waves. This diagnostic can be applied both to observations of solar and stellar flares and to future observations of non-flaring loops at high resolution.
\end{abstract}

\keywords{Sun: activity --- Sun: corona --- Sun: flares  --- stars: flare --- stars: coronae}

\section{Introduction} 
\label{sec:intro}

One fundamental question about coronal heating is whether the energy released in coronal loops is gradual or impulsive \citep{Klimchuk2006a,Parnell2012a,Reale2014a}. There is increasing evidence in the time series that in active regions the heating might be highly irregular \citep[e.g.,][]{Sakamoto2008a,Sakamoto2009a,Vekstein2009a,Terzo2011a,Viall2011a,Viall2012a,Ugarte-Urra2014a,Tajfirouze2016a,Tajfirouze2016b}. There are different colors about this question, for instance whether the heat pulses are frequent or not with respect to the plasma cooling time \citep[e.g.,][]{Warren2011a}. This is very difficult to constrain because coronal loops are likely structured into thin strands, where the heating is released through a storm of small-scale pulses. The single heating episodes can be hardly resolved up to date. In the framework of intermittent heating, an important issue is the duration of the heat pulses, because it links directly to the basic mechanisms of magnetic energy conversion, e.g., reconnection. For instance, recent work has suggested that the pulses are preferentially short \citep[$\leq 1$ min,][]{Testa2014a,Tajfirouze2016a}.

Even in flares the duration of the pulses is difficult to diagnose because the efficient heat conduction thermalizes the whole flaring loop in few seconds, cancelling all heating signatures in the EUV and soft X-rays. Some diagnostics about the features of the heating come from the hard X-rays, which track the emission from non-thermal electron beams. However, it is debated whether the electron beams are entirely responsible for the flare heating or they co-exist with other mechanisms excited by fast magnetic reconnection, e.g., the dissipation of current sheets \citep[e.g.,][]{Battaglia2015a}.

Here we propose a way to diagnose how long is the heating release in brightening coronal loops. Coronal loops can be described as closed magnetic tubes where the plasma is confined, moves and transports energy along the field lines. If the heat pulse is short, strong pressure waves are triggered inside the magnetic flux tube. Since the temperature is very uniform along the loop because of the efficient thermal conduction, the pressure waves \aftr{manifest as} steep density fronts that slosh up and down along the  tube. These are purely hydrodynamic waves. The density fronts determine visible periodic fluctuations in the light curves, which may be detected. 
This kind of fluctuations has been detected in the solar non-flaring \citep{Harrison1987a,Wang2003a} and flaring \citep{Nakariakov2009a} corona and in stellar coronal flares \citep{Mitra-Kraev2005a,Welsh2006a,Pandey2009a,Lopez-Santiago2016a}, and generally interpreted in terms of MHD harmonic modes.

In the following we investigate the details and conditions for the presence of these fluctuations through hydrodynamic loop modeling.

\section{The loop model} 
\label{sec:model}

We model the evolution of plasma confined in a closed semicircular magnetic flux tube, a coronal loop. The model has been extensively applied to study both coronal transients, such as solar and stellar flaring loops, nanoflaring strands, and loop brightenings \citep{Peres1987a,Reale1988a,Reale2000a,Reale2004a,Reale2005b,Reale2007b,Guarrasi2010a,Reale2012a,Reale2012b,Tajfirouze2016b}. Here the magnetic field has only the role to confine the plasma with no direct interaction with it, and the plasma can be described as a neutral fluid with pure hydrodynamics. Since the plasma flows and transports energy along the field lines only, in our model we can use one coordinate that follows the lines. 

We model the evolution of the loop plasma subject to an impulsive heating. We assume that the heating is deposited symmetrically with respect to the loop apex. 

We solve the 1D time-dependent hydrodynamic equations for a compressible plasma as described in \cite{Peres1982a}, including the effect of gravity (for a curved flux tube), thermal conduction, radiative losses for an optically thin plasma, compression viscosity. An external heating input consists of two contributions. One is steady and balances all the initial losses everywhere along the loop, keeping the unperturbed initial atmosphere at equilibrium. The other is time-dependent and describes the transient input that ignites the flux tube. It is a separate function of time and space. As \aftr{basic} time dependence we use a pulse function that is 1 for the duration of the pulse and 0 at any other time. \aftr{We made some tests also with a triangular pulse, with equal rise and decrease times.} As space dependence we assume a heating uniformly distributed along the loop. As a check for generality, we test also a twin heat pulse deposited at both loop footpoints.

The initial atmosphere is the usual hydrostatic corona linked to the chromosphere through a steep transition. The chromosphere is taken from standard models \citep{Vernazza1981a}, and is kept at equilibrium by a temperature-dependent heating function \citep{Peres1982a}.

As reference conditions, we consider a loop half-length (the loop is symmetric) $L = 25000$ km; we made tests also for a loop 4 times longer ($L= 10^5$ km). The initial loop atmosphere is relatively cool and tenuous. For the shorter loop, the density at the apex is $n_0 \approx 10^8$ cm$^{-3}$, the pressure $p_0 \approx 0.02$ dyn cm$^{-2}$, and the temperature $T_0 \approx 6 \times 10^5$ K, kept constant by a coronal heating rate $H_0 = 2.4 \times 10^{-5}$ erg cm$^{-3}$ s$^{-1}$. For testing, we have  considered also an initially warmer and denser atmosphere ($n_0 \approx 1.5 \times 10^9$ cm$^{-3}$, $p_0 \approx 0.8$ dyn cm$^{-2}$, $T_0 \approx 1.9 \times 10^6$ K, $H_0 = 1.2 \times 10^{-3}$ erg cm$^{-3}$ s$^{-1}$).

In this equilibrium atmosphere, we inject a heat pulse. As a reference case, we consider such a pulse intensity to bring the loop temporarily to a temperature above 10 MK. This mimics a heating typical both of medium class flares and of the hottest strands in a non-flaring multi-fibril active region loop \citep[e.g.][]{Reale2011a,Testa2012c,Tajfirouze2016a}. For the shorter loop, the heating rate is $H = 1$ erg cm$^{-3}$ s$^{-1}$ ($H = 0.05$ erg cm$^{-3}$ s$^{-1}$ for the longer one), which brings the loop to a maximum temperature $T \approx 16$ MK. For generality, we have also considered a 100 times lower heating rate ($H = 0.01$ erg cm$^{-3}$ s$^{-1}$), which brings the loop to a temperature about 4 MK.

In this study, the key parameter is the pulse duration. We have found that the reference time scale to trigger the plasma sloshing is the loop return (isothermal) sound crossing time:

\begin{equation}
\tau_s = \frac{2L}{c_s} \approx 50 \frac{L_9}{\sqrt{T_7}}
\label{eq:tau_s}
\end{equation}
where $c_s$ is the sound speed, $L_9$ is the loop half-length in units of $10^9$ cm and $T_7$ is the maximum loop temperature in units of $10^7$ K. For a maximum temperature of 16 MK, we obtain $\tau_s \sim 100$ s. We bracket this time scale with two pulse durations, $t_H = 60$ s and $t_H = 120$ s. We test the 4$\times$ longer loop with a short pulse duration of $t_H = 300$ s, and the 4$\times$ cooler loop with $t_H = 180$ s.

With these settings, the hydrodynamic equations have been solved by means of the Palermo-Harvard numerical code, with adaptive mesh refinement \citep{Peres1982a,Betta1997a}.

\section{The simulations} 
\label{sec:simul}

In the following we will describe in detail the results for two basic simulations, those with the reference loop half-length, (uniformly distributed) heating rate, initial atmosphere and with the pulse durations $t_H = 60$ s and $t_H = 120$ s.

\begin{figure}[!ht]               
\centering
\subfigure[]
 {\includegraphics[width=6cm]{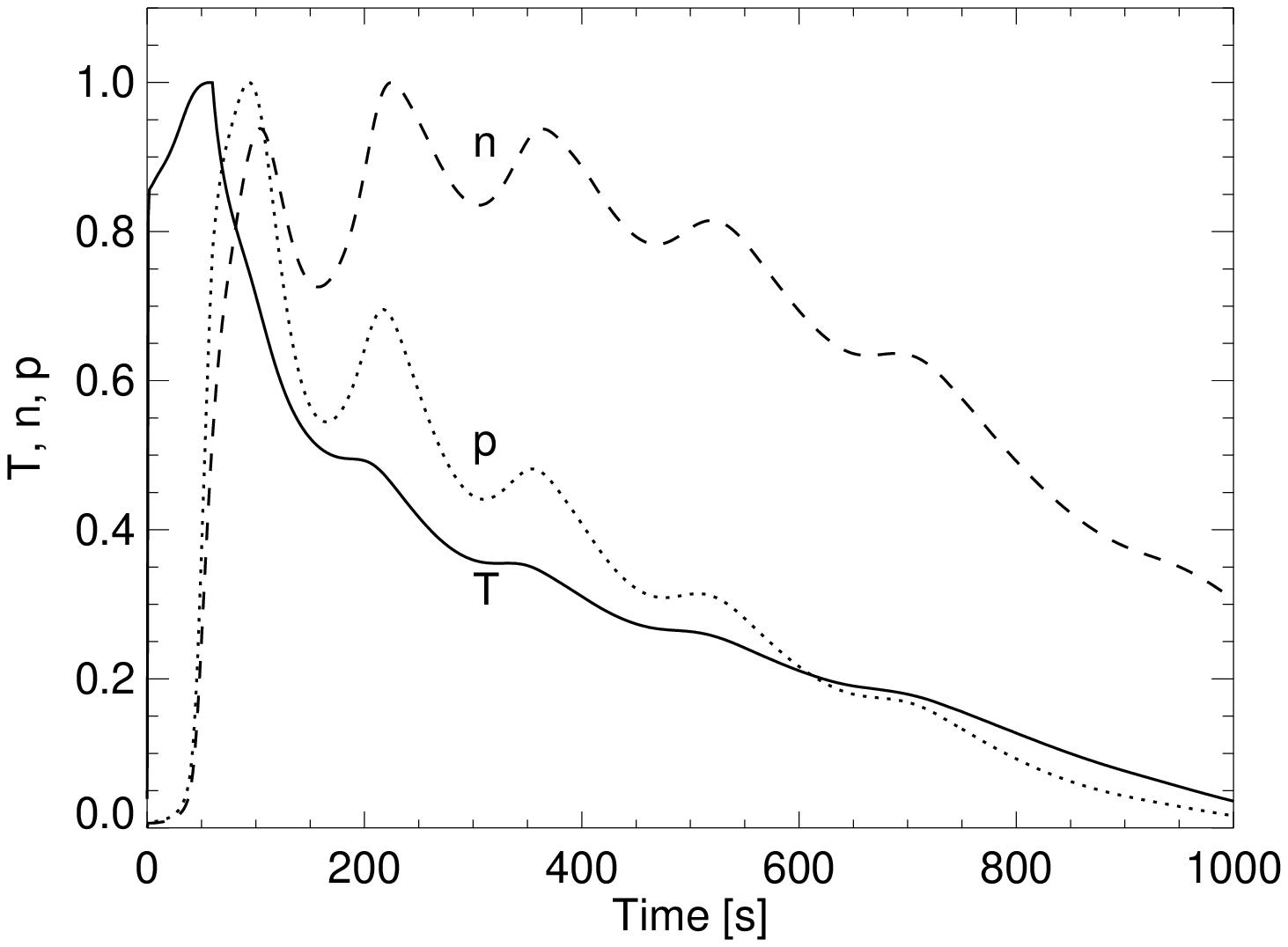}}
\subfigure[]
 {\includegraphics[width=6cm]{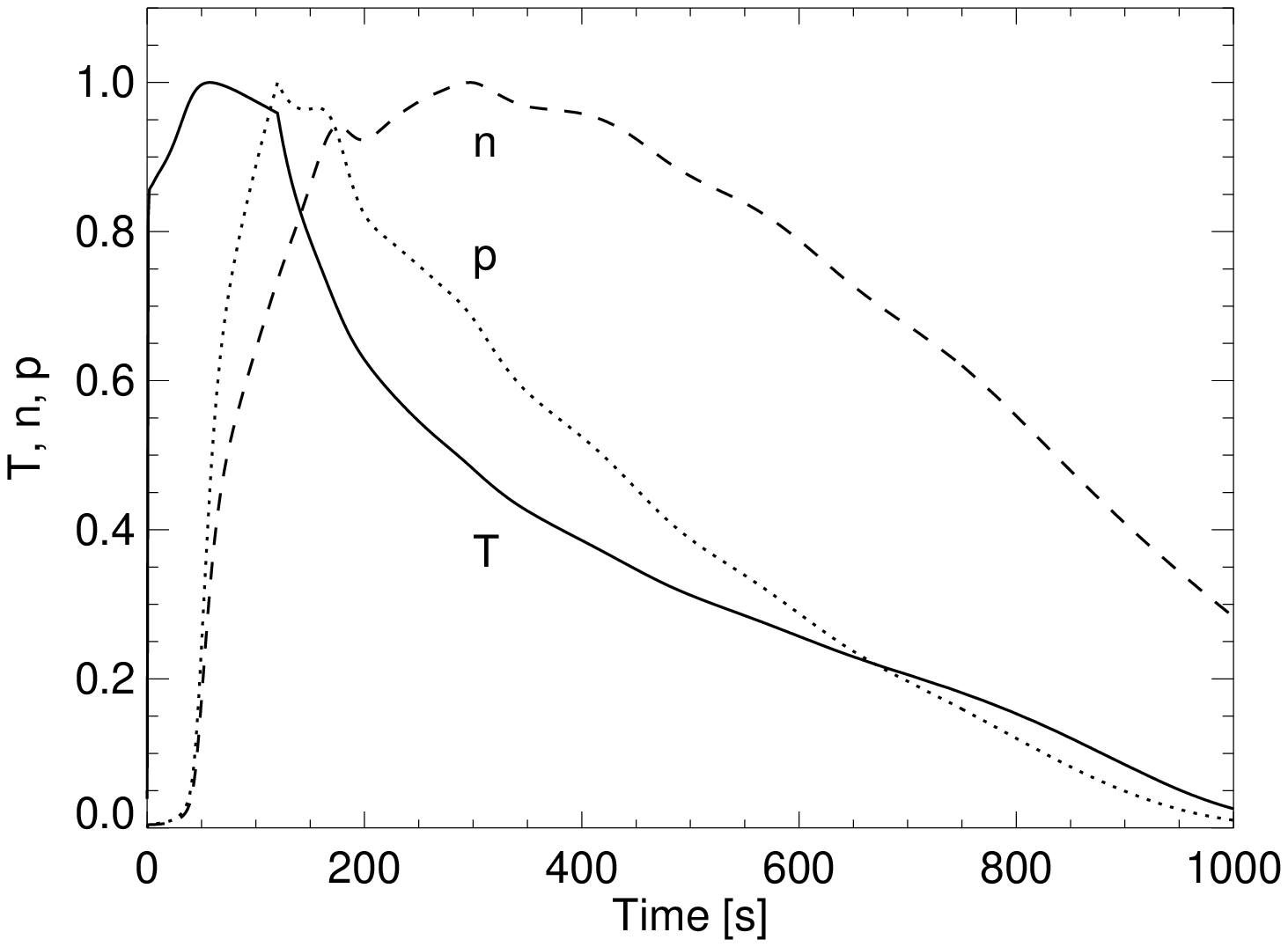}}
  \caption{\small Evolution of the temperature (solid), pressure (dotted) and density (dashed) at the loop apex for the pulse duration $t_H = 60$ s (a) and $t_H = 120$ s (b).  The values are normalized to 16.4 MK, 51 dyn cm$^{-2}$, and $1.67 \times 10^{10}$ cm$^{-3}$, respectively, for panel (a) and 16.4 MK, 78 dyn cm$^{-2}$, and $2.44 \times 10^{10}$ cm$^{-3}$ for panel (b).}
\label{fig:t_evol}
\end{figure}

Since the heat pulse is much stronger than the equilibrium heating rate, the temperature  rises to the flare value in a few seconds and throughout the loop because of the very efficient thermal conduction. The pressure increases as well and the overpressure drives an explosive expansion of the chromosphere upwards in the corona (chromospheric evaporation). After the heat pulse stops, the plasma rapidly cools down, again because of thermal conduction, while the density still increases for a while because the dynamic time scale is longer than the conduction time scale. Eventually, when the conduction cooling is replaced by the radiative cooling \citep[e.g.,][]{Cargill1994a}, the plasma begins to drain and the density decreases. \aftr{This evolution is well-known from standard loop modeling \citep[e.g.,][]{Bradshaw2006a,Reale2008a,Cargill2013a,Bradshaw2015a}.} Here we focus on the different modulation of this evolution determined by the different pulse duration.  

Figure~\ref{fig:t_evol} shows the evolution of the temperature, pressure and density at the loop apex for the two different pulse durations. For the first 60 s the evolution is of course identical. Then it differentiates, and the temperature drops for the short pulse duration and stays high longer in the other case. During the cooling the evolution is radically different because strong oscillations are present for the short pulse duration and not for the long one. These modulate the evolution and are best visible in the pressure and the density. From Fourier analysis, the period of the most powerful component is $\sim 150$ s, slightly increases with time, and 5 full oscillations are clearly visible in the time range to 1000 s (shown in the figure), after which the density has dropped to $\sim 1/3$ of the maximum value. Such clear oscillations are instead not visible for $t_H = 120$ s. The oscillations have a $\sim 20$\% amplitude in the density, and are damped with time. 

\begin{figure}[!ht]               
\centering
\subfigure[]
 {\includegraphics[width=6cm]{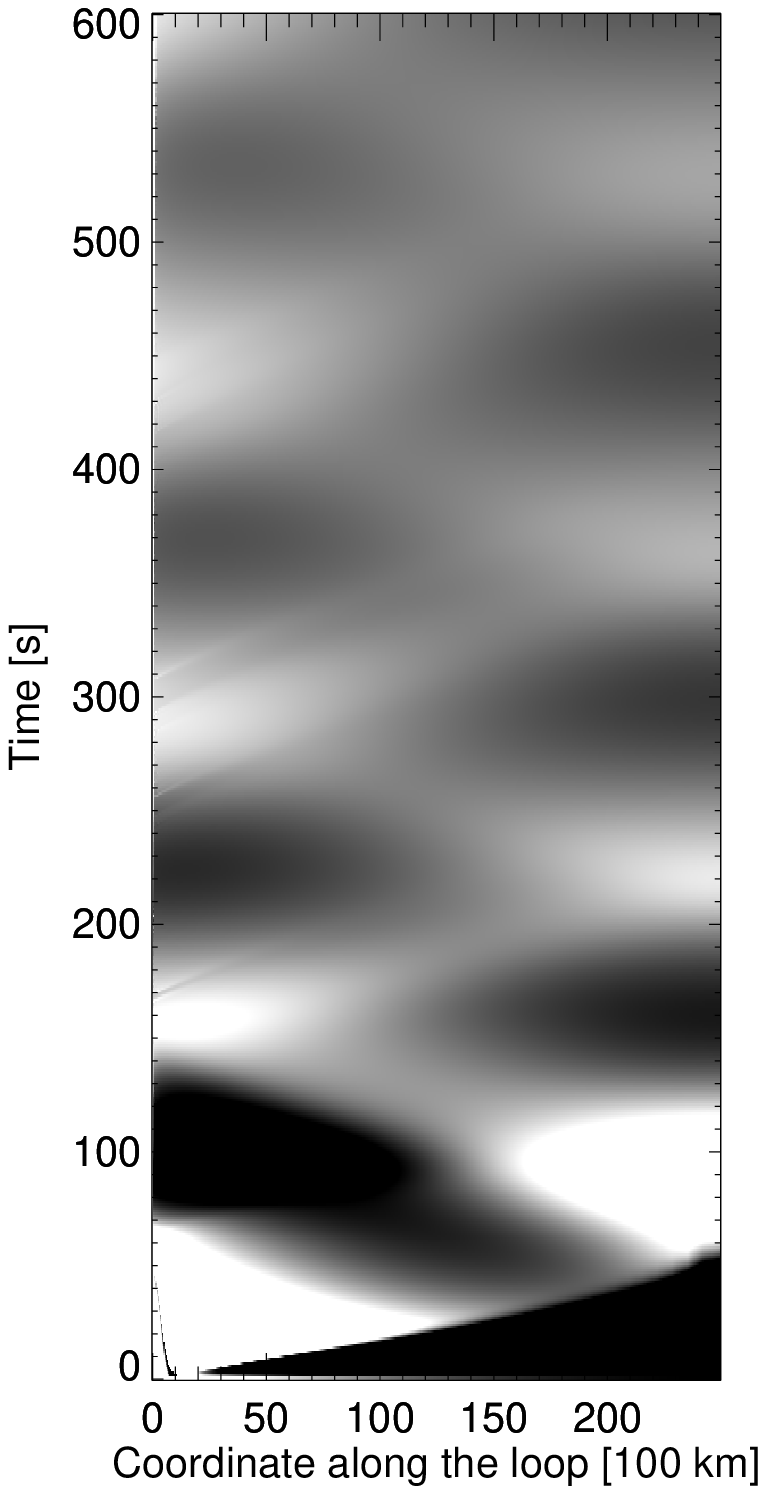}}
\subfigure[]
 {\includegraphics[width=6cm]{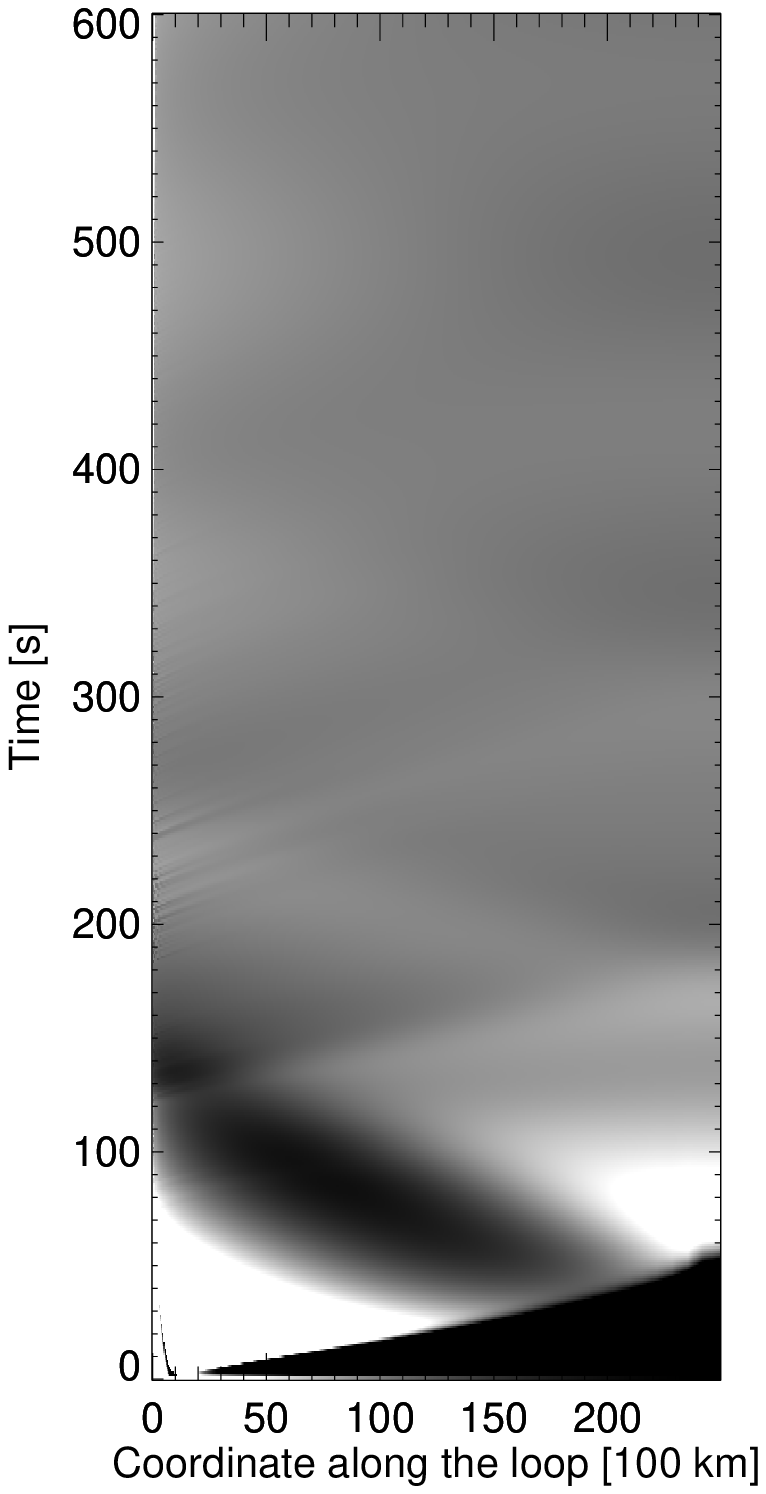}}
  \caption{\small Evolution of the pressure along (half of) the loop for the pulse duration $t_H = 60$ s (a) and $t_H = 120$ s (b).  To emphasize the presence of moving fronts, the pressure is normalized to the mean coronal value at each time, and the grey scale is between 80\% and 120\% of the mean value. }
\label{fig:p_front}
\end{figure}

They are the signature (at the apex) of a twin density (or pressure) front that sloshes up and down between the footpoint and the apex of the loop. Figure~\ref{fig:p_front}a ($t_H = 60$ s) shows this front very clearly in the form of bright zigzagging bands.
The front is more contrasted at the beginning, because of the sudden heating from a cool condition and then it slowly fades. For $t_H = 120$ s (Fig.~\ref{fig:p_front}b) we only see the initial bright pressure front, but this is promptly damped already at the first way down, and is very faint afterwards. 

\begin{figure}[!ht]               
\centering
\subfigure[]
 {\includegraphics[width=6cm]{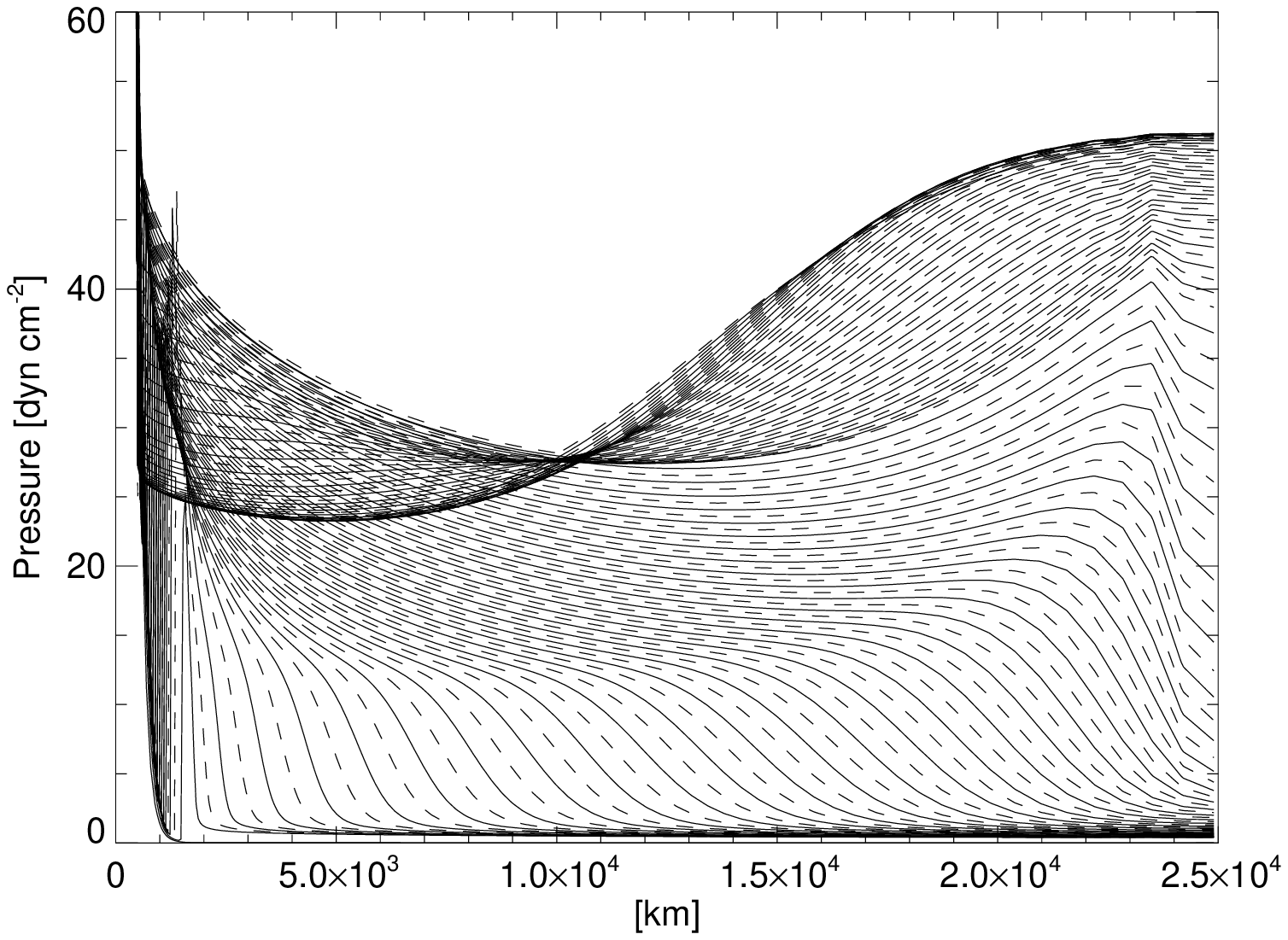}}
\subfigure[]
 {\includegraphics[width=6cm]{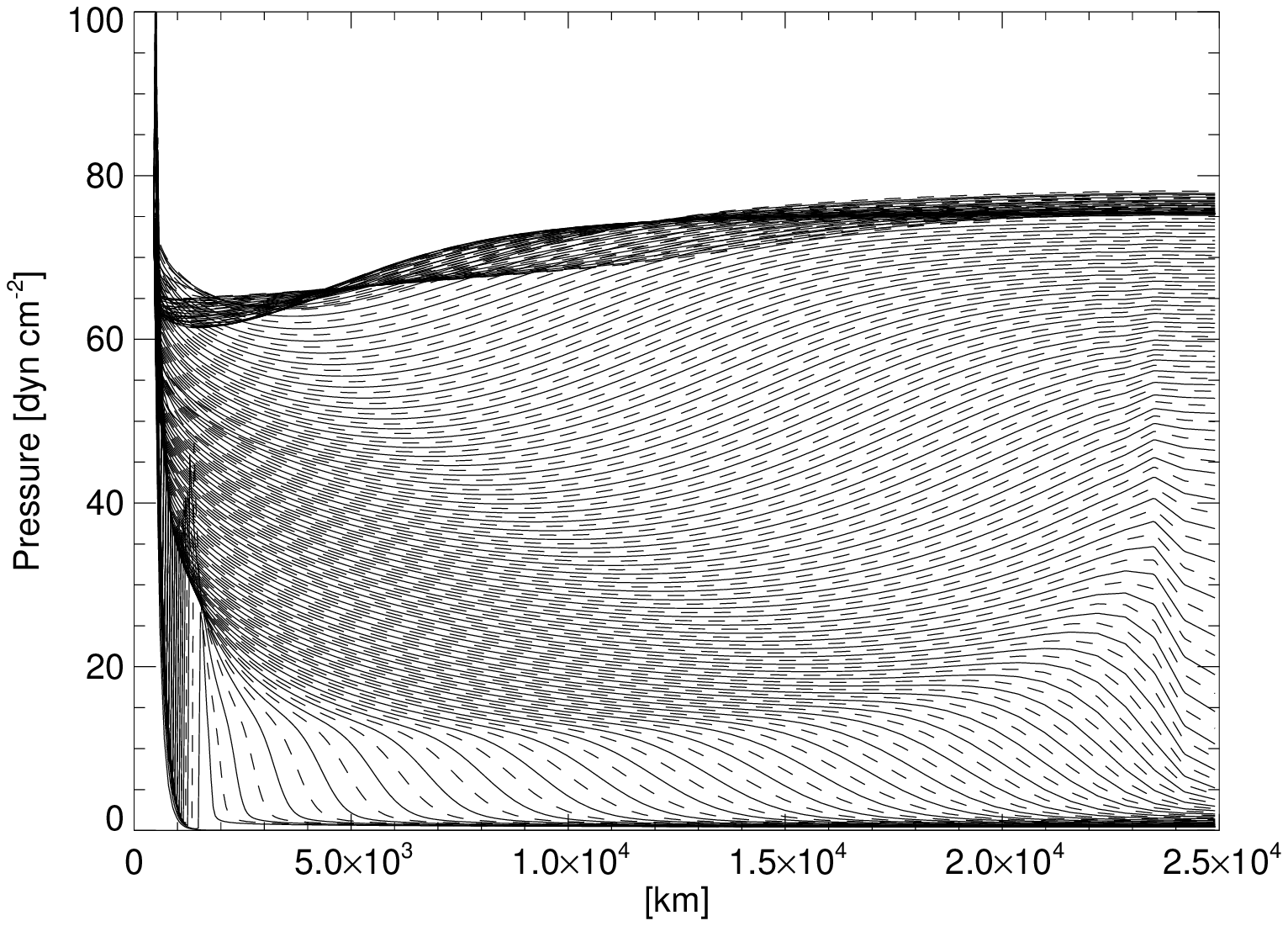}}
  \caption{\small Pressure profile along (half of) the loop for the pulse duration $t_H = 60$~s (a) and $t_H = 120$~s (b) in the first 100~s and 160~s, respectively, at time distances of 1 s (alternating solid and dashed lines). }
\label{fig:p_2d}
\end{figure}

The reasons for this different evolution can be understood from Fig.~\ref{fig:p_2d}. In both cases we see the initial steep evaporation front coming up (rightwards in the figure) from the chromosphere. After this, for $t_H = 60$ s (Fig.~\ref{fig:p_2d}a), the pressure continues to increase at the loop apex (right side), because plasma accumulates there. However, the pressure does not increase as well low in the loop (left side) and as soon as the the heat pulse stops, it even suddenly drops. This depression makes the upper steep pressure front travel backwards (downwards) along the loop (from right to left in the figure). The pressure drops as the temperature drops, because of conduction cooling. From Fig.~\ref{fig:t_evol}, the temperature decreases by more than 30\% in less than a minute,  much less than $\tau_s$ necessary to equalize the pressure. For $t_H = 120$ s (Fig.~\ref{fig:p_2d}b), the heat pulse lasts long enough to sustain the plasma and to equalize the pressure along the whole loop. Therefore, the critical process is whether the pressure equilibrium is reached or not along the loop, which explains why the sound crossing time is the key parameter.

We checked that we find very similar results, i.e., significant plasma sloshing when the heat pulse duration is shorter than $\tau_s$, for the longer loop, for the denser initial atmosphere, for heat pulses deposited at the loop footpoints, for the weaker heat pulse. \aftr{For the triangular pulse, we find that sloshing is still present for  120 s pulse duration, and cancelled for one twice as long. Therefore, if the pulse is more gradual than a square one, the threshold (\ref{eq:tau_s}) still holds using an ``equivalent duration'' that shrinks the triangular pulse to an equivalent square pulse. } Overall, this is a very general finding. It is worthwhile to remark that: a) the pulse duration should be compared to $\tau_s$ estimated at the temperature maximum \aftr{(or more correctly at the temperature of the pressure maximum)}, because the sloshing is triggered by the initial pressure imbalance; b) the period of the oscillations is essentially the time taken by the pressure/density front to travel back and forth along the loop; since during this time the plasma has considerably cooled down, the period is quite longer than $\tau_s$. For instance, in Fig.~\ref{fig:t_evol} the initial period of 150~s corresponds to a sound crossing time for a temperature of $\sim 7$~MK. The period (slightly) increases with time because the wave speed decreases in the cooling medium.

\begin{figure}[!ht]               
\centering
\subfigure[]
 {\includegraphics[width=6cm]{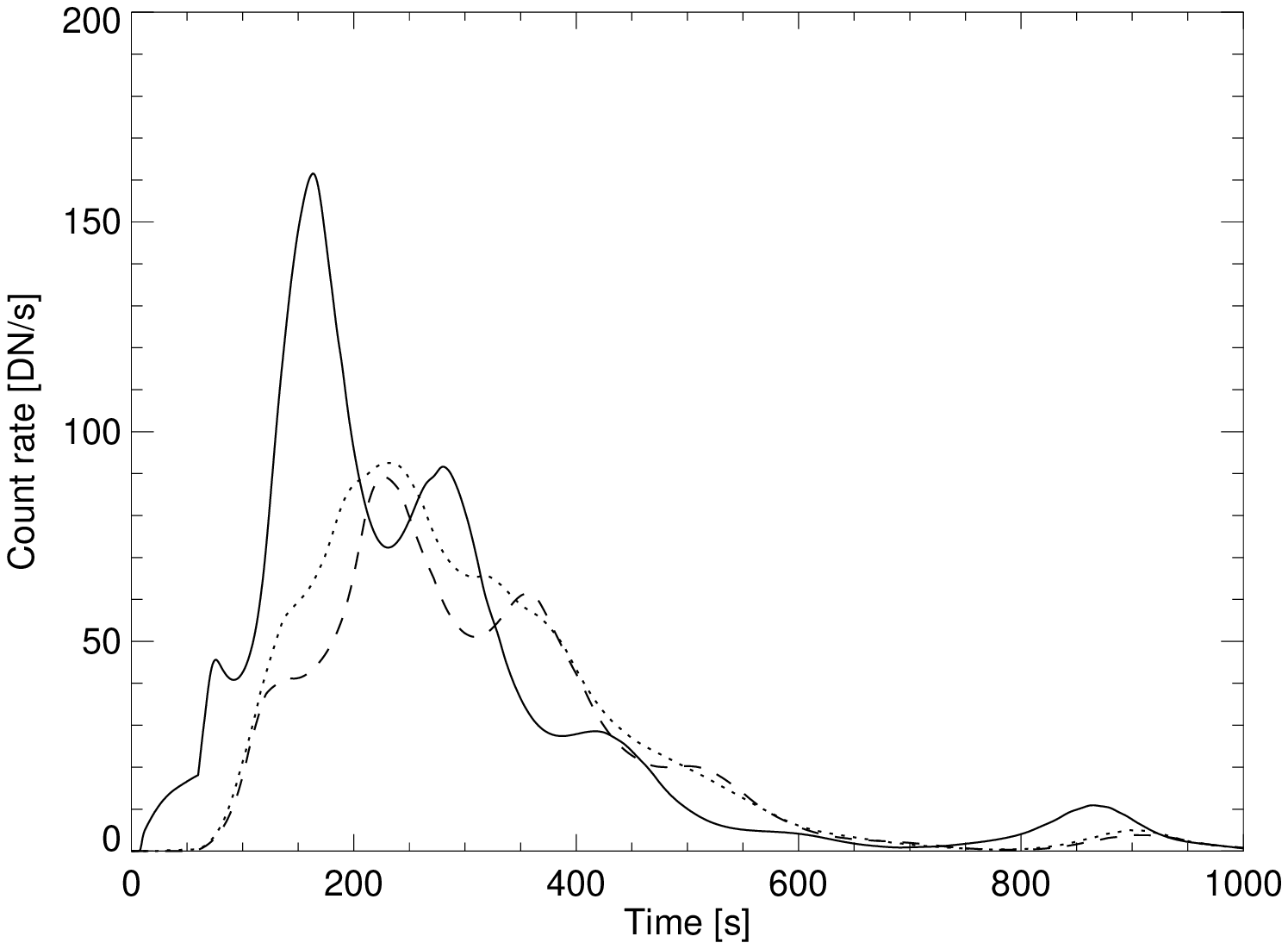}}
\subfigure[]
 {\includegraphics[width=6cm]{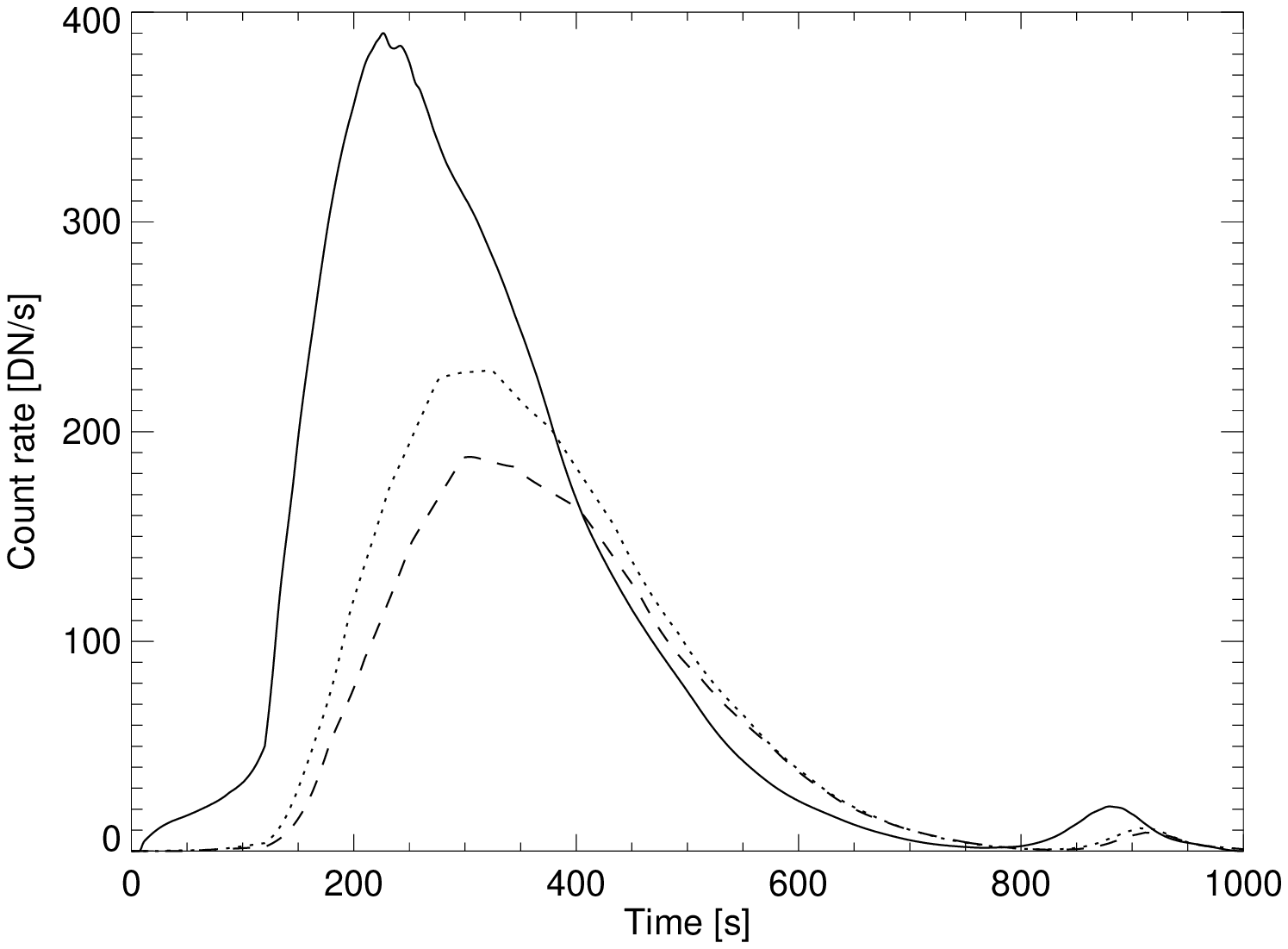}}
  \caption{\small Light curves in the SDO/AIA 94 A channel synthesized from the simulations with the pulse duration $t_H = 60$ s (a) and $t_H = 120$ s (b) at three different positions along the loop, i.e. footpoint (solid), apex (dashed) and in the middle between them (dotted). }
\label{fig:lc}
\end{figure}

As a final step, we explore whether this effect would be visible in the observations. We synthesize the emission as it would be observed in the 94~\AA\ channel by the Atmospheric Imaging Assembly \citep[AIA][]{Lemen2012a} on board the Solar Dynamics Observatory \citep[SDO][]{Pesnell2012a}. We choose this channel because it is sensitive to the emission from 5-10 MK plasma, but the results are general for any band or spectral line that is best sensitive to the emitting plasma. We assume that the plasma is optically thin and use the standard channel response function taken from the SolarSoftWare. \aftr{We also assume ionization equilibrium \citep{Reale2008a}, which holds on relatively long evolution times as those considered here. } The cross-section area is 1 pixel ($0.6" \times 0.6"$). Figure~\ref{fig:lc} shows the light curves obtained from segments 1000 km long at three different positions along the loop for both heat pulse durations.

Although overall we see the fast rise and slower decay typical of flare events, the light curves from the short pulse simulation are strikingly different from those of the long pulse one. They are very irregular, with periodic modulations that resemble those in the density (Fig.~\ref{fig:t_evol}). The fluctuations are even amplified because the emission depends on the square of the density. Larger fluctuations are present at the footpoint (the emission is more intense there because the density is higher due to gravity stratification). The light curves from the long pulse simulation are instead smooth and do not show significant fluctuations. The small late bump ($t \sim 900$~s) -- also present in the short pulse light curves -- is due to the second sensitivity peak around 1 MK in the 94~\AA\ channel \citep{Foster2011a}. Figure~\ref{fig:lc} shows that the fluctuations driven by the plasma sloshing are detectable in the light curves.

\section{Discussion}
\label{sec:discus}

We have shown that short heat pulses can excite large amplitude wavefronts of plasma confined in coronal loops. The critical time scale is the return sound crossing time (or the sound crossing time along the entire loop length) at the temperature peak. If the pulse duration is shorter than this time scale, then there is not enough time to equalize the pressure in the initial transient, and plasma sloshing is triggered back and forth between the apex and the footpoints. Since the efficient thermal conduction keeps the temperature very uniform along the loop, the pressure fronts are mostly density fronts, and determine strong fluctuations in the emission that can be detected in the light curves taken in the appropriate bands. 

We remark that we are modeling plasma flowing freely along the flux tube and that there is no direct interaction with the magnetic field, except for confinement and channelling. The excited waves are therefore purely hydrodynamic waves in a compressible plasma, different from low-order MHD modes, such as sausage or kink modes. 

The assumption of closed loops symmetric with respect to the apex makes the model evolution particularly clear and well-behaved. This scenario is \aftr{an acceptable simplification} because we might expect that magnetic reconnection triggers heat pulses deposited symmetrically at both loop footpoints. Also if the heat pulses are spread in the coronal part of the loop, the efficient thermal conduction would level out the temperature along the whole loop. Twin density fronts would anyway arise from the chromosphere at both footpoints and with a very small time difference, and they would hit against each other high in the loop determining the initial accumulation that triggers the sloshing. This might occur not exactly at the loop apex, so, if the evolution is not totally symmetric, we might expect more irregular quasi-periodic patterns.
\aftr{In addition, loops that are not symmetric could have very different gravitational stratifications in each leg, leading to more irregular patterns.}

The amplitude of the density waves is large and even larger in the plasma emission, because of the dependence on the square of the density. These are not standing waves nor acoustic harmonic oscillations inside the loop \citep{Selwa2005a}, and they have been customarily found in previous loop modeling \citep[e.g.,][]{Reale2012a,Bradshaw2013a}.


We might expect to detect them easily in the light curves, whenever present. At variance from typical magnetoacoustic waves, their period is relatively large, minutes (or more for longer loops), and therefore easy to identify. They may be best detected if the heating is released almost all at once across a loop, to have a coherent evolution, as in proper flares. Also, the large amplitude makes them different from typical MHD waves.

Another point is interesting to remark. A general decay time has been found for plasma flaring in single loops \citep{Serio1991a,Reale2007b,Reale2014a}:

\begin{equation}
\tau_d = 120 \frac{L_9}{\sqrt{T_7}}
\label{eq:tau_d}
\end{equation}
This decay time scales exactly as the sound crossing time~\ref{eq:tau_s}, i.e., the period of the waves scales as the decay time of the flare. Since the decay is typically the longest part of a flare, the implication is that, whatever the flare duration, any flare light curve will contain a similar number of major oscillations (not many, typically around 5).

In the end, we propose that periodic oscillations detected in the light curves of solar and stellar flares are often due to plasma sloshing as modelled in the present study and that their presence depends on the duration of the flare heating related to the flaring loop length (whereas the dependence on the temperature is relatively weak). Thus, this becomes a new way to identify pulsed heating and to constrain its duration. This does not seem to be so frequent in solar flares, probably because the length of the loops that brighten initially is often quite small. In spite of the smaller signal to noise, it is instead more frequent in stellar flares which can occur in giant magnetic channels \citep{Lopez-Santiago2016a}.

We can extend this result to flares at any scale, and in particular to small scales (nanoflares). We could expect to detect oscillations in light curves from non-flaring coronal loops in active regions, as observed, for instance, by SDO/AIA. This is not typically the case \citep[e.g.][]{Sakamoto2008a,Sakamoto2009a,Viall2011a,Tajfirouze2016a,Tajfirouze2016b}. but short heat pulses may still be present in the framework of multi-stranded pulse-heated loops. \cite{Tajfirouze2016a} find that short pulses match better the observed light curves and \cite{Tajfirouze2016b} show that, even if there are strong oscillations in the single light curves, they are washed out when they are mixed up along the line of sight across a loop with a multitude of independently heated strands.
We might hope to detect such oscillations even in non-flaring loops with appropriate resolution of next generation instruments.

\acknowledgments
The author thanks P. Cargill, J. Lopez-Santiago and the anonymous referee for suggestions, and acknowledges support from italian Ministero dell’Universit\`a e Ricerca.


\end{document}